 \documentclass[10pt]{article}

\usepackage{listings}
\usepackage{xcolor}
\usepackage{graphicx}
 \usepackage{url} 
 \usepackage{float} 
 \usepackage{fancyvrb}

\begin{document}

\title{LLM Agents for Generating Microservice-based Applications:\\
{\normalsize How Complex is Your Specification?}}
%
%
\author{Daniel M. Yellin\\
daniel.yellin@post.runi.ac.il}

\maketitle


 \begin{abstract}
In this paper we evaluate the capabilities of LLM Agents in generating code for real-world problems. Specifically, we explore code synthesis for {\it microservice-based applications}, 
a widely used architectural pattern for building applications. We define a standard template for specifying 
these applications, and we propose a metric for scoring the difficulty of a specification. The higher the score, the more difficult it is to generate code for the 
specification. Our experimental results show that agents using strong LLMs (like GPT-3o-mini) do
fairly well on medium difficulty specifications but do poorly on those of higher difficulty levels. This is due to more intricate business logic, a greater use of 
external services, database integration and inclusion of non-functional capabilities such as authentication. We analyzed the errors in LLM-synthesized code
and report on the key challenges LLM Agents face in generating code for these specifications. Finally, we show that using a fine-grained approach to code generation
improves the correctness of the generated code.   
 \end{abstract}

\maketitle

\section{Introduction}
The rapid maturation of Large Language Models (LLMs) for code generation has widely influenced software development,
with LLM Agents now routinely used for software engineering tasks including code generation. Benchmarks have become the
de facto standard for measuring an LLM Agent's performance. Early benchmarks, such as HumanEval\cite{DBLP:journals/corr/abs-2107-03374} 
and MBPP\cite{DBLP:journals/corr/abs-2108-07732} determine how
well LLM Agents perform in generating a single function of increasing algorithmic complexity. State-of-the-art models
(including reasoning models such as GPT-o1, GPT-o3 and DeepSeek-R1) show impressive results against these benchmarks, with results approaching 100\%
for HumanEval~\cite{BestLLMs2025}.

 However, these benchmarks shed little light on how well these models can perform on real-world problems where the
 complexity  is of a different nature. In real software development, complexity expresses itself not only
 in the computation (algorithm) of a particular component, but in the integration of different software components and
 in the distributed nature of modern systems; i.e., Cloud Native architectures~\cite{CloudNativeBookOReilly}. Newer benchmarks, such as 
 SWE-Bench~\cite{SWE-Bench:journals/corr/abs-2310-06770} have been developed to address realistic scenarios by 
 asking LLMs to fix a specific bugs in real GIT repos~\cite{zhang2023repocoder, liu2024repobench, 
Multimodal:yang2024swebenchmultimodal, Aleithan11024333} or to synthesize code to add a new feature/function to a 
repo~\cite{DevEval:li-etal-2024, CoderEval:conf/icse/YuSRZZMLLWX24, FEA-li-etal-2025}.

 The main challenge that GIT repo benchmarks poise to LLM code generation is the size of the context. Popular GIT repos
 often have hundreds of thousands LOC~\cite{SWE-Bench:journals/corr/abs-2310-06770} which is too large to fit into LLM context windows. (The
 average length of 500 real-world repositories is 1.1 million tokens\cite{li2024evocodebenchevolvingcodegeneration}.) 
Hence newer techniques, including agentic solutions, often include information retrieval techniques and focus on how to extract the right context to feed the 
LLM~\cite{NEURIPS2024_5a7c9475-SWE-agent, OpenHands:conf/iclr/0001LSXTZPSLSTL25, Agentless:journals/corr/abs-2407-01489}.

 There is an orthogonal problem in generating software for real-world problems that has received less attention; namely,
 how to \textit{specify} a complete solution to be generated. Real-world software is complex and requires
 far more descriptive text than required for fixing a GIT repo issue. How well do LLM Agents perform in this scenario?

 To make the problem concrete, we focus on using LLM agents to synthesis code for a core design pattern of modern systems,
 \textbf{Microservice-Based Applications} (\textbf{MSBA}s). These apps have the following characteristics:
\begin{itemize}
    \item \textbf{Microservices (MSs) }as the architectural pattern for system decomposition~\cite{MicroservicesBook}, usually deployed in
    containers such as Docker.

    \item \textbf{RESTful APIs }as the core communication mechanism~\cite{REST-Book:biehl2016restful}. 

    \item Use of \textbf{3rd party packages and external services} to easily incorporate existing functionality.
\end{itemize}

 The GitHub repo problem focuses on code generation where the code base is complex and cannot fit into the context window of the LLM. The MSBA problem, in contrast, focusses on code 
 generation from scratch, where the specification can easily fit into the context window of the LLM, but the challenge is in the complexity of the spec; i.e., the {\it intricacies} of the software 
 being developed,  not code size. For instance, SWE-Bench issues provided to an LLM to fix are 195 words on average\footnote{The max length for SWE-Bench issues is 4477 words.}~\cite{SWE-Bench:journals/corr/abs-2310-06770}. Our MSBA examples have descriptions of 1399 and 1905 words.  While these are not full-fledged
 production-level apps, they serve as a good testbed for our experiments.   As far as we know 
 little if any research has investigated using agentic software to generate MSBAs.   
 
 The purpose of this paper is to provide insight on how well LLM Agents  perform in this domain, analyze the types of errors they encounter, and show that the use of fine-grained 
 code generation enhances performance.  We also suggest a metric for rating the complexity of a specification.

\section{Challenges in generating MSBAs}
\label{sect-challenges}
Some of challenges in synthesizing MSBAs from LLM Agents are:

• \textbf{Integration}: reconciling type and naming mismatches between different frameworks and libraries. 

• \textbf{History and exceptions}: the response to a REST request can depend on the history of previous requests, and
depending on that history, a request could generate a normal result status code and return value, or one of many
exceptional (HTTP) status codes.

• \textbf{Non-functional considerations}: Orthogonal non-functional considerations, such as authentication or logging,
complicate code generation.

• \textbf{Dependencies}: A single MS can depend upon many other MSes and 3rd party services~\cite{DevEval:li-etal-2024}.  
This also requires understanding the right packages and versions to use.

• \textbf{Distribution}: MSBAs are inherently distributed systems and must deal with complexities inherent in such systems.

• \textbf{Specification}:  There is no standard for specifying an MSBA to an LLM.  Unlike simple functions, a docstring and a signature do not
suffice.  

On the other hand, MSBAs using REST services often follow a well-established pattern (see Appendix \ref{appendix-1-shot} for a small example) giving hope that LLM Agents can exploit
this pattern to generate correct code.   

The research questions we seek to answer are:
\begin{itemize}
\item How should one \textbf{specify} an MSBA to facilitate LLM code generation and how can one measure the complexity of a specification? This is discussed in Section \ref{MSBAspecification}.
\item How to \textbf{test} an MSBA? We address this in Section \ref{MSBAtesting}.
\item How well do LLM Agents \textbf{perform} in generating MSBAs? How does the complexity of the 
specification affect LLM performance? Our experimental results are given in Section \ref{section-experiments}.
\item Can a fine-grained approach to code generation can improve the accuracy of the generated code?  We show our results in Section \ref{section-fine-grained}. 
\end{itemize}
Sections \ref{section-related-work} discusses related work and Section \ref{section-conclusion} presents our conclusion.

\section{Specifying an MSBA}
\label{MSBAspecification}
To describe a microservice (MS) that is part of an MSBA you need to identify:
\begin{itemize}
    \item The {\bf resources} that are part of the interface. These are the “nouns” of an MSBA, the things that get operated  upon. For instance, 
    the resource /books is a collection of books, while /books/\{id\} is a specific book.
    \item The {\bf requests} (HTTP verbs) that can be issued on these resources.  The basic verbs are the CRUD requests POST, GET, PUT 
    and DELETE. Each request will specify additional details.   For instance, a GET request must specify its payload (parameters), 
    allowable query strings, 
    and details specific to HTTP and RESTful APIs, such as media types accepted and status codes to be returned.  The spec may also limit 
    whom is allowed to issue requests.  
    \item The {\bf computations} that must be performed and invariants that need to be maintained when handling a request.
    \item {\bf Dependencies} such as external services required to process requests and how to access their API keys.
\end{itemize}

\begin{figure}
\begin{Verbatim}[fontsize=\small] 
{"Name": "Cardholders microservice",
"Description": "Responsible for creating, retrieving, updating, and deleting 
     cardholder accounts. This service also provides the ability to retrieve 
     how much money in fines is owed by a specific cardholder.”
"Resources": "/cardholders, /cardholders/{id},/cardholders/fines/{id}. An id is 
     of type string. Each cardholder (/cardholders/{id}) consists of a JSON 
     document containing the field names ‘name’, ‘email’, and ‘id’. The values 
     of each of these is a JSON string. The resource /cardholders/fines/{id} 
     consists of a JSON document with a field ‘fineAmount’, whose value is 
     of type float, and a field ‘id’, of type string.",     
"REST requests": "GET and POST request on /cardholders. GET returns a 
     JSON array of all records in the resource. The GET request may specify a 
     query string of the form ‘field=value’. In this case, only those cardholder 
     records satisfying that constraint should be returned. The POST request 
     provides a JSON document containing the name and email and returns the 
     id for the newly created record.
     GET,PUT and DELETE requests on /cardholders/{id}. The GET request 
     returns a JSON record of the requested resource.  The PUT request provides 
     a JSON payload of the updated JSON resource’s fields and returns the id of 
     the updated resource. The DELETE request returns the status of the request 
     with no content. 
     GET request on cardholders/fines/{id}.  The request returns the JSON 
     document representing this resource.",
"Additional details”: "The fine amount owed is computed as follows: the fine per 
     overdue book is equal to number-of-days-overdue * FINE_PER_DAY, where 
     FINE_PER_DAY equals $.50. The cardholder owes the sum of all fines for all 
     books he borrowed. To compute this amount, this service needs to invoke the 
     Borrows MS to retrieve information on books borrowed by this cardholder.",
"Deployment": "The Cardholders MS runs in a Docker container and listens on 
     external port 5001. Another MS can send it requests using the URL 
     http://cardholders:5001/cardholders."}
\end{Verbatim}
\caption{Cardholders microservice}
\label{cardholders-example}
\end{figure}

In Figure \ref{cardholders-example} we provide the specification of a Cardholders MS, part of a public library application we present in Section 
\ref{section-library-app}. 
“Description” provides the purpose of the MS.   “Resources” lists the REST resources that the service is responsible for.  “REST requests” defines the type of REST requests supported for each resource, and how resources are to be represented.  “Additional details” provides functional requirements for the service.  “Deployment” states deployment details; for instance,  that the micro-service will run in a Docker container, the port the MS listens on and the URL used to send this MS requests.   An alternative to this format would be to use OpenAPI (Swagger) 
specifications~\cite{OpenAPISpecificationv3.1.1} but we would need to add information necessary to implement the server logic, 3rd party dependencies, etc. 
The format we provide is more compact.

\subsection{Creating MSBA specifications}
To evaluate  LLM Agents on MSBAs, we created 8 MSes across 2 MSBAs. We created our own MSBAS for two reasons. First we want to make
sure that the LLMs have not been contaminated - have not seen these MSes before.   Second, we want to compare LLM's code
generation capabilities on MSBAs of increasing specification complexity, but still small enough to focus on core MSBA challenges and not
on extraneous issues.  We built the MSBA specifications using the following process:
\begin{enumerate}
    \item We wrote the \textbf{spec} of each microservice as part of a complete MSBA specification, and reviewed it with an LLM for completeness and consistency.
    \item We wrote a  \textbf{Ground Truth}  (\textbf{GT}) implementation for the spec.
    \item We wrote \textbf{unit test cases} based upon the  spec and used them to test the GT solution.
    \item When testing and finding an error, we checked the cause of the error. If it was a mistake in the GT
   implementation, we fixed the GT implementation. If the test case was not faithful to the spec, we corrected the test case.
   If the spec was ambiguous or incorrect, we fixed the spec.
    \item We repeated step 4 until the GT implementation passed all tests.
    \item We used the spec to generate preliminary code from an LLM and tested that code with the unit tests. We discovered a few
    more errors this way, and repeated until we were confident in the correctness of the GT implementation and unit test cases.
\end{enumerate}

The specifications and list of unit tests for each MSBA can be found online\footnote{
https://github.com/LLMs4code/LLMs4MSBA\label{footnote-online}}.

\subsection{The complexity of an MSBA specification}
\label{spec-compexity}
We measure the complexity of an MSBA in several ways:
\begin{enumerate}
    \item {\it Size}: the number of words in the MSBA specification. Although we generate each MS one at a time, 
    we provide the entire MSBA specification to the LLM because it often needs the entire spec to generate even a single MS correctly.
    \item {\it Resources}: the resource complexity.  A resource is either a primitive data type (int, string,...), a structure, where each field is a resource, or a 
    collection of resources.   If a resource is a single primitive it counts as a single {\it primitive resource}.  If it is a collection, it counts as a single
    {\it collection resource}.  If it is a structure with $k_1$ fields that are primitive or structures and $k_2$ fields that are collections, it counts as $k_1$ primitive and
    $k_2$ collection resources.  Structures or containers containing embedded structures or containers are recursively counted on their own.     
    We denote the resource complexity by
    X-Y where X is the number of primitive resources and Y is the number of collection resources in the application as defined by the spec\footnote{
    Sometimes a resource is computed dynamically - the data is not actually stored.  In that case we do count the resource.  If a resource does not
    represent data but is used to take action (e.g.,`/turn-light-on') it also is not counted at all.
    }.
    
    \item {\it Dependencies}: Let $M$ be a MS and $E$ be a service endpoint.  $M$ is {\it dependent on}
    $E$ if $M$ invokes $E$.   $E$ may be another MS in the application or it may be an external service.  We count the number of 
    dependencies for each MS in the MSBA, and the sum of these is the {\it dependency complexity} of the MSBA.  We use the GT code to 
    determine the number of dependencies. Note that if $M$ invokes
    $E$ from multiple places in the code, it is still counted as a single dependency.  
    \item {\it Packages} is the number of packages the MS requires.   We count the number of packages that the GT 
    implementation imports.
    
    \item {\it LLM-as-a-Judge} is the score, between 1 and 5 assigned by the LLM to the difficulty of implementing the spec.  
\end{enumerate}

The specification size indicates the number of details the LLM must contend with.   The number of dependencies and packages 
indicates the amount of external knowledge required by the LLM.  The number of resources indicates the data management complexity.
These metrics are not exact; its possible to have a large specification that is simple to implement and a small specification that is difficult to 
implement.
Similarly not all external dependencies are equal;  some express more intricate interfaces than others.   Nonetheless, these metrics give a rough estimate
on the complexity of an  MSBA spec and are easy to compute.   The LLM-as-a-Judge serves as a validation for the other metrics, and it also 
explains why it rated the spec with the given score, providing further insight into the complexity of the spec.  Additional metrics could be added, such as 
the number of conditional requirements/constraints in the specification, similar to measuring Cyclomatic Complexity in 
programs (see also \cite{li-etal-2024-control}).

\section{Testing MSBAs}
\label{MSBAtesting}
The result of a request to a MS is sensitive to the request history.  For instance, the outcome of a
GET request is dependent on the preceding POST, PUT and DELETE requests. One needs to test multiple
scenarios for a GET to determine whether it functions correctly.  Furthermore, the unit tests must also check
that the appropriate
exceptions are raised when warranted; e.g., when a GET request is issued on an item previously deleted, the status code
404 must be returned whereas if the request provides the wrong media type a status code 415 is returned. Developing adequate
unit tests can be time consuming  and tedious. To help simplify and make testing of REST APIs more consistent, we developed 
a REST API
testing framework, that consists of 4 modules.

The main module is (i)  \textit{Unit Test}, which is comprised of {\it common} unit tests and {\it MS-specific} issues the unit tests.  It relies on the (ii) \textit{HTTP Service
Controller} to format and invoke the requests according to the HTTP protocol.  The common
tests are those where the result depends on the pattern of the requests, without concern for their specific input
and output values. For instance, if we issue requests POST $X_1$, POST $X_2$, POST $X_3$, DELETE $X_2$
and GET, we expect the return value from the GET to be the JSON array containing $X_1$ and $X_3$ and nothing else. This is
true independent of the actual values of $X_i$. We rely on this insight to parameterize common unit tests. Parameter values,
expected returned values, and any other MS specific values needed to test the MS are defined in the (iii) \textit{Database module}. The
Unit Test module fetches the appropriate values from the DB when invoking a common unit test on a MS; for the example above, it would fetch
the values for $X_1$, $X_2$ and $X_3$ relevant for the specific MS.  

The common unit tests may still require some customization, and not every such test applies to all MSes. However, the
framework greatly simplifies building tests for a new MS. Likewise, we created a reusable parameterized (iv) \textit{assertion}
library that checks that return values satisfy assertions. For example, the assertion
\footnotesize
\texttt{assert-fields-contains-value(response: requests.Response,f:str,v)} 
\normalsize 
asserts that the response is a JSON array, each element in the array is a struct with field f, the field f is
itself a JSON array, and contains the value v.

In addition to common unit tests, we test MS-specific functionality. For example, we check that the Cardholders MS (Figure
\ref{cardholders-example}) correctly computes the fine for overdue books.
MS-specific tests need to be customized per MS.

Say that an MSBA is made up of 4 MSes: $M_1$, $M_2$, $M_3$, and $M_4$.  Let $M^{GT}_i$ be the GT implementation
of $M_i$ and $M^{A}_i$ be the implementation of $M_i$ generated by the LLM agent. To test $M^{A}_2$ we run the MSBA
using the implementations $M^{GT}_1$, $M^{A}_2$, $M^{GT}_3$, and $M^{GT}_4$ and then test the application.  In this
way we isolate the test failures due only to $M^{A}_2$. 

\section{Experiments}
\label{section-experiments}
\subsection{Methodology}
\begin{figure}[h]
    \centering
    \includegraphics{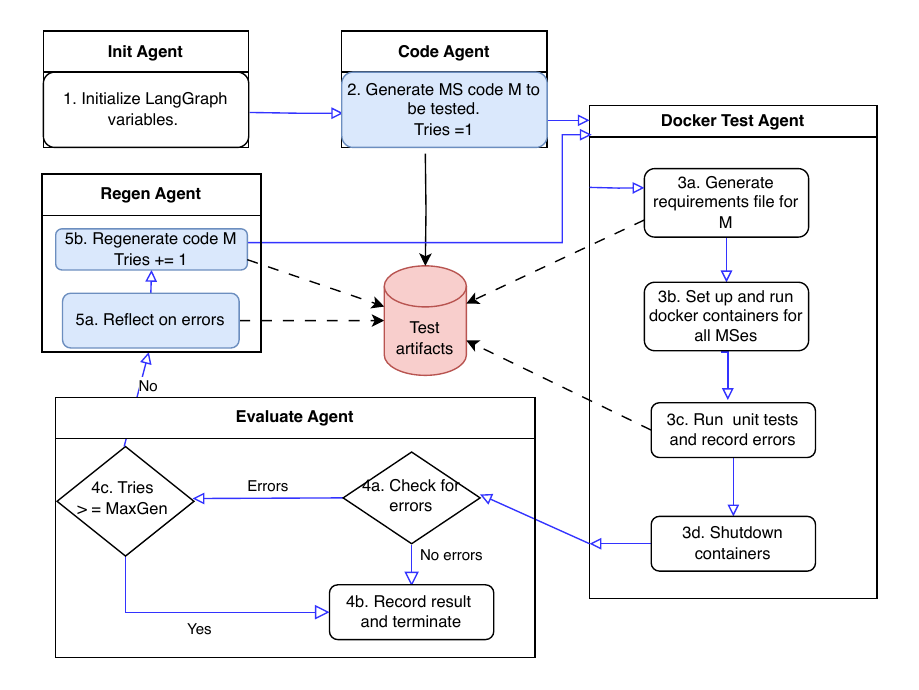}
    \caption{Workflow for generating and testing an MSBA.  The blue arrows indicate control flow.  The dotted black arrows indicate writing and reading data.  Colored boxes denote LLM interactions.}
    \label{workflow-diagram}
\end{figure}

We developed an agentic system built on top of LangGraph~\cite{Langgraph-url} and given in Figure~\ref{workflow-diagram} to 
generate and test the microservices (code) of an MSBA given its specification\footnote{
We refer to each node of the LangGraph as an agent for simplicity of discussion.  However, it is commonly accepted to refer to the
entire LangGraph as a single agent, and each node of the graph as a function that is part of the agent.  
}.
The Init Agent (1) initializes and setups the graph.  The Code Agent then (2) generates code for 
a MS, based upon the specification.   Once this code is generated, it passes control to the Docker Test Agent.   This agent first (3a) generates a requirements
file for the generated code.  It then (3b) sets up a Docker container to execute the LLM generated code $M$ and Docker containers to execute all the other
MSes using the GT code for each of them.   When these containers are up and running the agent (3c) executes the test script on the application and logs
any run-time or test errors.   Finally it (3d) shuts down all the containers and frees resources before passing control to the Evaluate Agent.  

The Evaluate Agent (4a) checks the logs produced by the Docker Test Agent.  If there were no errors, then (4b) it records the result and terminates.   Otherwise
(4c) it checks if the system has reached the maximum number of code-generation trials.  If yes, (4b) it records the result and terminates.  Otherwise 
it passes control to the Regen Agent, who (5a) asks the LLM to reflect~\cite{Reflexion:conf/nips/ShinnCGNY23} on the errors and produce a plan to fix the
generated code to prevent the top 5 errors.   
Because of the context window limitations of LLMs, and because the runtime may produce a very long trace of each error, we use a script to remove
parts of the error messages that are not useful to debug the issue.    For some LLMs, we need to further limit the
size of the error messages to fit into the context window.    
(5b) The original code together with the result of the reflection is then provided to the LLM, which is instructed to generate code that repairs the MS.    
Control then passes back to the Docker Test Agent (3a) to test the regenerated code. The entire process is repeated for each MS.  
The prompts used by the agents are given in Appendix~\ref{appendix-prompts}.   

In summary, the agentic system automates the entire process from code generation to testing to repair by invoking multiple agents.  It uses LangGraph to 
orchestrate the flow, runs the generated code safely in Docker containers, and uses reflection to improve results.  

\subsection{Installing packages}
\label{section-package-problems}
The Docker Test Agent runs the generated code in a Docker container.  The Dockerfile for that container contains a command to install dependencies in the requirements.txt file using {\it pip}.    The agent generates the requirements.txt file by scanning the generated code for 
import statements\footnote{In our prompt we asked the LLM to document 
which packages to install.  But we found that it would make mistakes and it was a better solution to install packages based upon the import statements.   
}. This was problematic as {\it pip} would fail when installing
a package that comes with the standard python distribution (such as {\it uuid} or {\it json}). To get around this problem, we omitted packages 
included in the Python standard library.  

In the restaurant application 
(Section \ref{section-restaurant}), the generated code for the authentication MS would always import the {\it jwt} package, but it required the installation of
{\it pyJWT}.   (To its credit, the LLM would note this in a comment.)   For this and other similar issues, our code substitutes the package name in the import statement with the correct
package.   Nonetheless, as we show below and documented by other research~\cite{Package-Hallucinations-1:journals/corr/abs-2406-10279, 
robustimports:latendresse2025, GitChameleon:journals/corr/abs-2411-05830}, choosing the right package and version is still problematic.

\subsection{The Public Library Application}
\label{section-library-app}
Our first MSBA is a public library application comprised of 4 MSs: Cardholders,
Books, Borrows, and Logs.  The Cardholders
service was given in Figure \ref{cardholders-example}.  Books is responsible for managing the books in the library and uses the
Google Books API service (see https://developers.google.com/books/docs/v1/using) to supply additional information about a book.  
Borrows is responsible for tracking books on loan, and Logs keeps a record of each
book that is borrowed from the library.   The specification of these MSes can be found on-line (see footnote~\ref{footnote-online}).

We test this application using 3 different LLMs to power the Code and Regen Agents: GPT-3.5-16K, GPT-4o-mini, and GPT-o3-mini.   For each 
LLM we performed two sets of experiments, one using a $0$-shot prompt, and one using a $1$-shot prompt~\cite{Few-Shot-Learners:NEURIPS2020_1457c0d6}.  The $1$-shot prompt, 
given in Appendix \ref{appendix-1-shot}, provides example code using Flask and MongoDB for a small MS.
For each LLM and each MS of the MSBA we repeat each experiment 15 times.  For GPT-3.5 and GPT-4o-mini we perform 5 experiments
for each temperature $0.0$, $0.3$, and $0.5$.  GPT-o3-mini did not support different temperature settings
and hence all experiments used the default temperature.  
The Docker Test Agent tests the MSBA with unit tests for each MS.   There are 15 unit tests for Cardholders, 14 for 
Books and Borrows, and 7 for Logs.   The latter has fewer tests 
because Logs only allows external actors to perform GET requests; they cannot create, delete or modify logs on their own. 
For each experiment, if the LLM generated code does not pass all the unit tests, the Docker Regen Agent prompts the LLM to reflect on the nature of the errors 
and then prompts it to regenerate code, as explained above.   We only do this
once (MaxGen = $2$ in Figure \ref{workflow-diagram}).  We refer to the initial generated code as V0 and the regenerated code as V1.

\subsubsection{Results}
Tables \ref{figure-gpt35-0-shot} and \ref{figure-gpt35-1-shot}  shows the results of our experiments on GPT-3.5 with the $0$-shot and $1$-shot prompts
respectively.  The $1^{st}$ column 
is the name of the MS.  As not all generated code was executable, the $2^{nd}$ column lists the fraction of the 15 
experiments that generated executable V0 code permitting execution of the unit tests.  The $3^{rd}$ column lists the number of experiments that
generated executable V1 code.   The $4^{th}$ and $5^{th}$ columns present 
the percent of unit tests that passed using V0 and V1 code respectively, averaged over all experiments that produced executable code\footnote{
We do not consider unexecutable generated code in this percentage.      
We wanted our results to focus on how often executable code was functionally correct.
}.  The $6^{th}$ column is of
the form $n_1$-$n_2/t$, where $n_1$ ($n_2$) is the highest number of unit tests passed in any experiment for V0 (V1) and 
$t$ is the total number of unit tests for that MS.
The last column is the percent of experiments in which the number of unit tests passed by V1
was greater than the number of unit tests passed by V0, considering only those experiments in which V1 was executed.\begin{table}
\begin{tabular}{| c | c | c |  c | c | l | c|}
\hline 
Name &  \# V0              &  \# V1         &  \% V0       &  \% V1    & Highest       & \% V1  \\
           & testable          &  testable     & passed     & passed   & \# passed   & improved  \\
\hline 
Cardholders  &  13/15  &  13             & 26.4     & 50.2       & \# 7-8/15    & 76.9  \\
\hline 
Books           &  15/15   &  15             &  38.1    & 37.1       & \# 7-7/14    & 13.3 \\ 
\hline 
Borrows        &  15/15  &  15             &  28.1     & 32.4        & \# 7-7/14   & 33.3 \\  
\hline 
Logs             &  15/15  &  15             & 30.5     & 26.7        & \# 4-4/7       & 26.7 \\
\hline 
\end{tabular}
\caption{Public library MSBA results using GPT-3.5 with $0$-shot prompt}
\label{figure-gpt35-0-shot}
\end{table}
   
\begin{table}
\begin{tabular}{| c | c | c |  c | c | l | c | }
\hline 
Name &  \# V0              &  \# V1         &  \% V0       &  \% V1    & Highest      & \% V1  \\
          & testable           &  testable     & passed     & passed   & \# passed   & improved  \\
\hline 
Cardholders  &  13/15  &  13    & 59.3     & 78.4      & \# 13-13/15      & 61.5  \\ 
\hline 
Books           &  15/15  &  15     &  61.9     & 66.2     & \# 13-13/14      & 40.0 \\ 
\hline 
Borrows        &  15/15  &  15     &  87.2    & 72.4      & \# 13-14/14      & 6.7 \\ 
\hline 
Logs             &  15/15  &  8       &  83.8     & 62.5     & \# 7-6/7            & 12.5 \\   
\hline 
\end{tabular}
\caption{Public library MSBA results using GPT-3.5 with 1-shot prompt}
\label{figure-gpt35-1-shot}
\end{table}

Tables  \ref{figure-gpt-o3-mini-0-shot} and \ref{figure-gpto3-mini-1-shot} present the results for GPT-o3-mini using the $0$-shot and
1-shot prompts.
\begin{table}
\begin{tabular}{| c | c | c |  c | c | l | c|}
\hline 
Name &  \# V0              &  \# V1         &  \% V0       &  \% V1    & Highest       & \% V1  \\
           & testable          &  testable     & passed     & passed   & \# passed    & improved  \\
\hline 
Cardholders  &  13/15  &  5      &  94.4    & 0.0      & \# 15-0/15         & 0.0  \\ 
\hline 
Books           &  13/15  &  12    &  83.5    & 16.7     & \# 14-14/14      & 16.7 \\ 
\hline 
Borrows        &  15/15  &  10    &  85.2    & 20.0      & \# 14-14/14     & 20.0 \\ 
\hline
Logs             &  14/15  &  4      &  90.8     & 17.9       & \# 7-5/7       & 25.0   \\   
\hline 
\end{tabular}
\caption{Public library MSBA results using GPT-o3-mini with $0$-shot prompt}
\label{figure-gpt-o3-mini-0-shot}
\end{table}
If V0 code passes all unit tests, V1 code would not be generated.   Hence sometimes there were very few V1 samples generated.   
If these samples have a major flaw (like uncommented text in the code that caused the entire MS to produce runtime errors), it resulted in the 
$0\%$ passing value in the $5^{th}$ column.
\begin{table}
\begin{tabular}{| c | c | c |  c | c | l | c |}
\hline
Name &  \# V0              &  \# V1         &  \% V0       &  \% V1    & Highest       & \% V1  \\
           & testable          &  testable     & passed     & passed   & \# passed    & improved  \\
\hline
Cardholders  &  13/15  &  4    &  89.2    & 0.0      & \# 15-0/15         & 0.0  \\ 
\hline
Books           &  15/15  &  14   &  81.4   & 21.4      & \# 14-14/14     & 21.4  \\ 
\hline
Borrows        &  15/15 &  13    &  87.6    & 30.8       & \# 14-14/14   & 26.7  \\ 
\hline
Logs             &  15/15  &  5     &  90.5     & 0.0         & \# 7-0/7        & 0.0   \\   
\hline
\end{tabular}
\caption{Public library MSBA results using GPT-o3-mini with 1-shot prompt}
\label{figure-gpto3-mini-1-shot}
\end{table}

We omit the results for GPT-4o-mini due to space considerations.  
In general it performs better than GPT-3.5 but not as good as GPT-o3-mini.  See Appendix  \ref{appendix-temperature}.

\subsubsection{Discussion}
Looking at column 3 of Table \ref{figure-gpt35-0-shot} we see that GPT-3.5 performs  
poorly with the $0$-shot prompt, never reaching 40\% average test success.   In contrast, in Table \ref{figure-gpt-o3-mini-0-shot} we
see that GPT-o3-mini performs quite strongly, 
averaging between 83.5 and 94.4\%.  Keep in mind that this is the average number of unit test cases passed;  it does not mean that
94.4\% of the instances pass {\it all} test cases.   While GPT-3.5's performance increases upon reflection (V1 code), it still lags significantly behind
GPT-o3-mini. Furthermore, from column 6 of Table~\ref{figure-gpt-o3-mini-0-shot} we see that some GPT-3o-mini V0 samples for each service pass all tests.
This affirms the strength of GPT-o3-mini in generating code.  

A second observation, based upon columns 5 and 7 of these tables, is that reflection is effective for GPT-3.5 but much less so for GPT-o3-mini.
 We attribute
this to two factors.  First, GPT-o3-mini was often successful in passing all unit tests using V0, and hence there were fewer V1 experiments.
Second, even when V0 did not pass all tests, it had a high success rate, so it created a higher bar for V1 to improve upon.   

A third observation, based upon column 4 of Tables \ref{figure-gpt35-1-shot} and \ref{figure-gpto3-mini-1-shot}, is
that providing an example (1-shot prompt) 
significantly improves performance of GPT-3.5 but mostly degrades performance with GPT-o3-mini.   This may be the case with reasoning
models like GPT-o3-mini\footnote{For instance, the creators of DeepSeek -R1 report
that ``When evaluating DeepSeek-R1, we observe that it is sensitive
to prompts. Few-shot prompting consistently degrades its performance. Therefore, we
recommend users directly describe the problem and specify the output format using a
zero-shot setting for optimal results."~\cite{DeepSeek-R1:guo2025deepseek}  See also~\cite{wang2024advancedlanguagemodelseliminate}.
}, and especially in our case, where performance is high without the example, and the example can cause
overfitting.   

Looking at the generated code, we find that GPT-3.5 often makes mistakes in the basic REST patterns and 
requires the example to nudge it to implement the REST APIs as specified.   In contrast, GPT-o3-mini understands this pattern well 
and does not require much assistance.  

A general comment is warranted on these statistics; namely, not all unit tests are equal. Some unit tests may require complicated
logic while others check very simple REST patterns that do not even require knowledge of the MS.  For instance, one of the unit tests checks
that a POST request with a non-JSON payload fails. Hence while percentage of unit tests passed still gives us an overall idea on the strength
of the LLM, we require a deeper analysis examining which tests pass or fail to obtain additional insight (see Section~\ref{sect-restaurant-results-discussion}).   

With respect to temperature, while the $0.3$ setting often performs best for V0 and $0.5$ for V1, the differences in performance between different temperatures is usually
less than $10\%$. See Appendix \ref{appendix-temperature} for details.     

\subsection{The Restaurant Application}
\label{section-restaurant}
The 2\textsuperscript{nd} application is a restaurant nutrition MSBA comprised of 4 MSs, Authentication,
Dishes, Profile, and Ratings.  Authentication  is responsible for authenticating users, including
registration, login, and token validation.  
Dishes is responsible for creating, retrieving, updating, and deleting dishes offered by the restaurant.   Only the 
restaurant owner (a predefined role) is allowed privileged operations such as POST, while general users 
are only allowed to perform GET requests. This service
invokes the 3rd party API Ninjas nutritional service, available at \textit{https://api-ninjas.com/api/nutrition},
to provide nutritional information on the dish.    
Profiles  is responsible for creating, retrieving, updating, and deleting user profiles. Each user profile
contains a nutritional diet for that user, and allows the user to retrieve only those dishes that satisfy his/her diet.   
Ratings is responsible for maintaining ratings of restaurant dishes and computing the average rating for a dish.  
All users can rate a dish but a user is only allowed to rate a dish once.   

Given the poor results of GPT-3.5 on the Library app, we use only GPT-4o-mini and GPT-o3-mini for our experiments on this MSBA. For similar
reasons,  we only experiment with a $0$-shot prompt.   For GPT-4o-mini we performed 20 experiments for each MS.  
For the ratings MS,
8 of the experiments were at temperature $0.0$, 7 were at temperature $0.3$ and 5 were at temperature $0.5$.   For all other microservices, there were
10 experiments at temperature $0.0$, and 5 experiments at each of the temperatures $0.3$ and $0.5$.
For GPT-o3-mini we performed 15 experiments for each MS. As mentioned above 
GPT-o3-mini did not support different temperatures so we used the default temperature. 

Each experiment consists of 9, 15, 10 and 8 unit tests for the Authentication, Dishes, Profile, and Ratings MSes respectively.  
As above, for each experiment, if the LLM generated code
does not pass all unit tests, the Regen Agent reflects on the nature of the errors and then regenerates code.   
 
\subsection{Results}
Tables \ref{figure-restaurant-gpt-4o-mini-0-shot} and  \ref{figure-restaurant-gpt-o3-mini-0-shot} give the results of testing on GPT-4o-mini 
and GPT-o3-mini LLMs.  We see that their performance degrades significantly on the restaurant MSBA compared to the library MSBA.
 
 \begin{table}
\begin{tabular}{| c | c | c |  c | c | l | c | }
\hline
Name &  \# V0              &  \# V1         &  \% V0       &  \% V1    & Highest       & \% V1  \\
           & testable          &  testable     & passed     & passed   & \# passed   & improved  \\
\hline
Authentication  & 0/20     & 20           &  0              & 57.7        & \# 0-6/9        & --    \\
\hline
Dishes             &  20/20  & 20            &  15.3        & 23.7         & \# 5-7/15     &  55.0   \\ 
\hline
Profile              &  20/20  & 19           & 3.5           & 8.4       & \# 1-1/10     & 47.3     \\ 
\hline
Ratings            &  20/20  & 20          & 1.3          & 0.6         & \# 1-1/8       & 5.0   \\   
\hline
\end{tabular}
\caption{Restaurant MSBA results using GPT-4o-mini with $0$-shot prompt}
\label{figure-restaurant-gpt-4o-mini-0-shot}
\end{table}

 \begin{table}
\begin{tabular}{| c | c | c |  c | c | l | c | }
\hline
Name &  \# V0              &  \# V1         &  \%V0       &  \% V1    & Highest       & \% V1  \\
           & testable          &  testable     & passed     & passed   & \# passed   & improved  \\
\hline
Authentication  & 13/15  & 11         &  65.8        & 79.3        & \# 8-9/9      & 81.8    \\
\hline
Dishes             &  15/15  & 15        &  23.6        & 25.8         & \# 5-6/15   &  20.0   \\ 
\hline
Profile              &  15/15  & 15        & 14.0         & 15.3         & \# 2-2/10   & 26.7    \\ 
\hline
Ratings            &  15/15  & 15       & 10.8          & 39.2         & \# 3-6/8    & 66.7   \\   
\hline
\end{tabular}
\caption{Restaurant MSBA results using GPT-o3-mini with $0$-shot prompt}
\label{figure-restaurant-gpt-o3-mini-0-shot}
\end{table}

\subsection{Description Complexity}
Before analyzing the results of these experiments, we first state the description complexity of the restaurant and library MSBAs, given
in Table \ref{MSBA-complexity}.  The description size is for the 0-shot prompt.  The 1-shot prompt adds $490$ words. 

The Restaurant MSBA has a description size about $.36$ larger than
the Library application, and has twice the number of dependencies, although the package complexity is the same.  The resource complexity of the
Restaurant application is slightly larger.  This at least partially
explains the difference in LLM performance between the library and restaurant MSBAs.  
 \begin{table}
 \footnotesize
\begin{tabular}{| c | c | c |  c | c | c | }
\hline
Name & Description            & \# Dependencies    & Avg \# Packages  & Primitive-Coll   & LLM assigned \\
           & size  (words)         &                                & per MS                 &  Resources      & difficulty \\
\hline
Library          & 1399  &    3     & 4             &  22-4  & 3  \\
\hline
Restaurant   &  1905  &   6      & 4             & 25-5   & 4 \\
\hline
\end{tabular}
\caption{Description complexity of MSBAs}
\label{MSBA-complexity}
\end{table}


The LLM-as-a-Judge score, provided by Claude Sonnet 4, confirms this conclusion.   It assigns the library application a difficulty of 3 (out of 5) while the
restaurant application received a complexity score of 4.   It says that the library app has ``moderate complexity" and that it is not overly
complex because of the ``Clear, well-defined requirements", ``Standard CRUD operations", ``Common technology stack", ``Simple data models", and 
``Limited business logic".   Factors that add complexity include ``Inter-service communication",``External API integration", ``Business rule enforcement", ``Date calculations", ``Error handling", and ``Distributed data consistency". 

For the restaurant MSBA, the LLM Judge provides the following ``High Complexity Elements" contributing to the 4 rating: ``Multi-service Architecture", 
``Authentication \& Authorization", ``External API Integration", ``Complex Business Logic", ``Database Operations", ``Error Handling",  and ``Query Processing". It 
did not give it a 5 rating because of ``Well-defined specifications with clear requirements", ``Standard technologies (Python, MongoDB, REST APIs)",
``No real-time processing or complex distributed systems patterns" and ``No advanced security beyond basic token auth".
 In Appendix \ref{appendix-LLMasJudge} we provide the  LLM-as-a-Judge prompt and the comments provided by the LLM to explain its score for the 
 restaurant application. The comments for the library app can be found on-line (see footnote~\ref{footnote-online}).
 
\subsection{Discussion}
\label{sect-restaurant-results-discussion}
From Table \ref{figure-restaurant-gpt-4o-mini-0-shot} we see that GPT-4o-mini does extremely poorly on the restaurant application, and while 
GPT-o3-mini (Table \ref{figure-restaurant-gpt-o3-mini-0-shot}) does better, it still never performs above $25\%$ passing unit tests with V0 code
except for the Authentication MS, where it achieves $65.8\%$.   We also note that neither V0 nor V1 code ever achieves a perfect score, with the
exception of the Authentication service.  
In 6 out of the 11 V1 testable implementations, it passed all  $9$ unit tests.  

To get a deeper understanding of the cause of the errors, we examined the run-time error-logs, the GPT-o3-mini reflection results, 
as well as the generated code.  Among the persistent problems are the following:
\begin{itemize}

\item {\bf Incorrect package usage}.   For instance, even though we installed {\it pyJWT} (see Section \ref{section-package-problems} ), the authentication code
would often use this package incorrectly.  Along similar lines, Ratings would consistently import
the {\it flask\_jwt\_extended} package.   A typical error message was the following:

\footnotesize
\verb+ **Error**:`ImportError: cannot import name `jwt_optional' from `flask_jwt_extended'+
\normalsize

We tried multiple versions of the {\it flask\_jwt\_extended} package but none corrected the problem, 
or would result in other runtime errors. In short, incorrect package usage
is often a fatal problem in the generated code~\cite{Package-Hallucinations-1:journals/corr/abs-2406-10279, robustimports:latendresse2025}.

\item {\bf Superficial understanding of APIs}.  This is best illustrated by the Dishes MS using {\it https://api-ninjas.com/api/nutrition}.   The spec purposely changes the
names of the nutritional elements used in the MS from those provided by the API.   Furthermore, for some nutritional elements, such as protein, the API 
returns the string  {\it ``Only available for premium subscribers"}.   These elements are not required by the app, but the code must know that not every element in the
array will be numeric or it throws an exception.  This level of API understanding was beyond the LLM.   To investigate further, we developed a new prompt 
that explicitly states the details of the API.   Although using this prompt the generated code was able to address these API complexities, 
the test results remained largely the same due to other issues, like those enumerated below.   Finding appropriate APIs and correctly implementing them is
another recognized LLM issue~\cite{API-Bank:conf/emnlp/LiZ000YLHL23, Autofeedback:journals/corr/abs-2410-06943, APIproficiency:conf/kdd/0002GKEFTS24, APIinteractions:journals/corr/abs-2409-11703}.

\item {\bf Lacking or incorrect datatype conversions}. Surprisingly, the generated Dishes implementation has difficulty converting the JSON payload to a string.
Its failure to do so results in a runtime error and the POST request fails. This has a cascading effect - since the POST fails, 
subsequent GET requests fail. This error is pervasive and contributes to the low passing rate for some of the services.   

A similar issue  is converting between REST APIs parameters, passed as a string, and the type required to use in an API invocation.   For example, 
a resource (e.g., a specific {\it dish}) is identified by its string ID in the REST API.   But MongoDB uses, by default, ObjectIDs to identify Mongo documents. Failure
to convert between these datatypes was a common error in the library application using GPT-3.5.  GPT-o3-mini generally performs this conversion. However there
are still times when it fails to do so due to subtleties of the Mongo API.   Specifically, the generated code:
\verb+dishes_collection.insert_one(dish_record)+
followed by 
\verb+jsonify(dish_record)+
causes a run-time error: \\
\verb+TypeError: Object of type ObjectId is not JSON serializable+.  This is because by default Mongo adds an ObjectId to the 
\verb+dish_record+.
One needs to remove this added field or convert it to a string before applying the \verb+jsonify()+ method.  Type inference is 
a problem for LLMs~\cite{Types:journals/corr/abs-2504-09246}.

\item {\bf Disregarding specification details}.   A good example of this problem is wrong return codes.   In the prompt we state ``If an authentication token is not provided or is invalid, then 401 (Unauthorized) is returned.''
However, the generated code returns status 400 (Bad request).  This same sort of error occurs for other return codes as well.  
Providing good error messages to the LLM on why a particular unit test fails will allow it to correct the problem. In this case, the LLM V1 
implementation was able to fix faulty return codes.   
\end{itemize}

\section{Fine-grained experiments}
\label{section-fine-grained}
We conjecture that using a fine-grained approach to generate complex code can improve the results.   
Specifically, instead of generating code for the entire service, the Code and Regen Agents  generate code for each request type (e.g., POST,
GET, ...) independently.   They then merge the generated code into the GT code 
for that service.  For example, if the Code Agent generates code for the POST request, it will then replace the POST request code in the GT file with the 
generated code but the rest of the GT code remains the same.  The Docker Test Agent still tests the entire application, but the results will be solely dependent
on the generated POST code.   The value of this approach is that it:
\begin{itemize}
\item Adopts a divide-and-conquer approach to code generation, allowing the LLM to focus on generating a smaller piece of code for each prompt.
\item Facilitates further understanding of the cause of the errors in the generated code, as the generated code is only for a specific request.   
\item Removes the possibility that an error in one request type (e.g., GET) is due to an error in a previous request of another type (e.g., POST).   
\end{itemize}
The GT code contains assumptions that all services must share, for instance, the name of the Mongo collection that stores service resources.    Furthermore, there
are cases where merging the generated code for a specific request
into the existing GT implementation can cause conflicts if care is not taken.   For this reason the prompts used by the Code and Regen agents
include guidance on how to generate the code: it specifies the name of the Mongo collection,  it requests that a wrapper be used for 
authentication, and it declares the name of packages that are available to import. Appendix~\ref{appendix-dishes-POST-prompt} provides the Code Agent
prompt for generating POST request code for the dishes service.   The prompts for other requests 
types and other services are similar.   

The results of the fine-grained experiments for the dishes and ratings services of the restaurant MSBA are given in figures 
\ref{figure-restaurant-dishes-gpt-o3-fine-grained} and \ref{figure-restaurant-ratings-gpt-o3-fine-grained} respectively.   
All experiments were performed using GPT-o3-mini. In these tables each row except for the last one presents the results
for generated code of the request type given in the $1^{st}$ column.   The $2^{nd}$ column lists the number of experiments performed.   The 
$3^{rd}$ column
lists the number of unit tests that explicitly test that request type in each experiment\footnote{For example, POST requests that explicitly test the POST 
function are included in these numbers.
POST requests that are used to set up a test on another request type, such as DELETE, are not included in these numbers.}
and the $4^{th}$ column and $5^{th}$ columns gives the number and percent of those tests that passed for V0 code.   The last two columns state the 
number and percent of those tests that passed for V1 code.   The last row presents the average number of passing unit tests for V0 and V1, averaged over
the rows above it.   For Table \ref{figure-restaurant-ratings-gpt-o3-fine-grained}, we consider it as if $100$ is written in the last column, since $100\%$ of the 
V0 tests passed and therefore no V1 code was generated.   

We first compare the fine-grained results on the dishes service to the coarse-grained results for that service.  We note that the latter results in $23.6\%$ and
$25.8\%$  for V0 and V1 code respectively (Table \ref{figure-restaurant-gpt-o3-mini-0-shot}, row 2): 
\begin{itemize}
\item The V0 results for the fine-grained generation of POST requests for the dishes service (Table \ref{figure-restaurant-dishes-gpt-o3-fine-grained}, row 1, col 5) 
yields approximately the same percent of V0 passing tests as for the dishes service in the coarse grained experiments.   

\item The V0 results for the fine-grained generation of the GET and DELETE requests (rows 2 and 4) show remarkable improvement in passing percent over the coarse-grained experiments.   
PUT (row 3)  is lower, but each experiment has a single PUT test and therefore one cannot overly rely on its results, given the scarcity of data.   

\item All the requests, including POST, show significantly better results on V1 code than the coarse-grained results.

\item The previous observations leads one to infer that (1) POST requests are harder to generate, and because the POST code often fails in the 
coarse-grained service generation, the subsequent requests will likewise fail. (2) Generating smaller pieces of code (at the request level) pays off.    
The average passing percent across all 4 request types for V0 and V1 code is $48.5\%$ and $77.8\%$ respectively 
(Table \ref{figure-restaurant-dishes-gpt-o3-fine-grained}, last row).  The is significantly better than $23.6\%$ and
$25.8\%$  for coarse-grained generation.   The improvement between V0 and V1 percents is also very much higher.   However, these observations are tempered by the 
knowledge that the fine-grained code is easier for the LLM to generate as we supply the LLM with information that is not given in the coarse-grained
prompts  (authentication code, import statements, and Mongo collection name.).   

\item A good example fine-grained generation improvement over coarse-grained generation
can be seen by the issue we cited  in Section \ref{sect-restaurant-results-discussion} ``Lacking
or incorrect datatype conversions''.   We noted there that the generated code \verb+dishes_collection.insert_one(dish_record)+
did not remove the Mongo \verb+_id+ before the jsonify call.  This was true for both the V0 and V1 code in the coarse-grained experiments.   While this was true 
for the V0 code in fine-grained experiments, for V1 code, $10/12$ times it corrected this error.  

\item  Generating fine-grained requests for the ratings service is less beneficial than for the dishes service.   Specifically, generating the POST V0 request code
fails in all experiments, as seen in Table \ref{figure-restaurant-ratings-gpt-o3-fine-grained} row 1.  Surprisingly, generating GET requests (row 2) likewise proves hard for the 
LLM in V0 code.   However, V0  PUT code is correct for every experiment (row 3).   

\item Nonetheless, there is still significant improvement in using fine-grained instead of coarse-grained generation for the ratings service.   The average passing 
percent across all 3 request types for V0 and V1 
code is $33.3\%$ and $63.2\%$ respectively (Table \ref{figure-restaurant-ratings-gpt-o3-fine-grained}, last row).  
The is much better than $10.8\%$ and $39.2\%$ percent for coarse-grained generation (Table \ref{figure-restaurant-gpt-o3-mini-0-shot}, row $4$, columns $4$ and $5$). 
\end{itemize}

\begin{table}
\begin{tabular}{| c | c | c |  c | c | l | c |}
\hline
Request  &  \# experiments & \# tests       & \# V0         & \% V0       & \# V1       & \% V1       \\
               &                           &                   & passed       & passed     & passed   & passed     \\
\hline
POST      & 12                      & 7               & 20/84          &  23.8         & 38/84       & 45.2         \\
\hline
GET        & 13                      &  5               & 51/65          &  78.5         & 58/65       & 89.2         \\ 
\hline
PUT        & 13                      &  1                & 0/13           & 0                & 10/13       & 76.9         \\ 
\hline
DELETE  & 12                    &  2                 & 22/24          & 91.7           & 4/4          & 100          \\   
\hline
Overall Avg  & 12                 &                    &                    & 48.5           &                & 77.8         \\   
\hline
\end{tabular}
\caption{Results for fine-grained generated code for the dishes restaurant service}
\label{figure-restaurant-dishes-gpt-o3-fine-grained}
\end{table}

\begin{table}
\begin{tabular}{| c | c | c |  c | c | l | c |}
\hline
Request  &  \# experiments & \# tests       & \# V0         & \% V0       & \# V1       & \% V1       \\
               &                           &                   & passed       & passed     & passed   & passed     \\
\hline
POST      & 12                      & 4                & 0/48           &  0              & 11/48       & 22.9         \\
\hline
GET        & 12                      &  3               & 0/36           &  0               & 24/36       & 66.7         \\ 
\hline
PUT        & 12                      &  1                & 12/12         & 100            & -              & -               \\ 
\hline
Overall Avg  &                      &                    &                    & 33.3           &                & 63.2         \\   
\hline
\end{tabular}
\caption{Results for fine-grained generated code for the ratings restaurant service}
\label{figure-restaurant-ratings-gpt-o3-fine-grained}
\end{table}

In summary, the results indicate that fine-grained request generation is a promising direction to pursue.   It also shows that additional focus needs to be given to generating
the POST request as it is the most difficult request type to generate, and if the POST code does not work properly, the other requests will also fail.   

\section{Related Work}
\label{section-related-work}
Extending LLM code generation to tackle realistic real-world problems for software engineers is an active area of research.   One trend in this direction is 
software engineering agents, who facilitate developers in their everyday tasks.   More directly related to our work is research in extending LLM code synthesis
to tackle more realistic and complex scenarios.
Much of this research is focussed on automatically correcting issues or extending
the functionality of a solution in a GitHub repository~\cite{SWE-Bench:journals/corr/abs-2310-06770, DevEval:li-etal-2024, Aleithan11024333, NEURIPS2024_5a7c9475-SWE-agent, FEA-li-etal-2025}. 
While the code repair or extension may be relatively minor, the difficulty is in locating the relevant context, which can be non-trivial 
given the size of the repository.    In contrast, our work focusses on taking a specification of an application that has multiple microservices and
generating code from scratch to realize the specification.   The challenge here is the complexity of the specification.   It requires generating code
that incorporates business logic, invokes the appropriate internal and external dependencies, incorporates non-functional considerations such as logging and
authentication, and handles a variety of inputs and multiple error conditions.

We found that for moderately difficult specifications GPT-o3-mini did quite well.  However, for difficult specifications it performed poorly.  
By examining the errors we found specific issues that it had trouble resolving.  Some of these problems have been identified in previous work.  
For instance, one issue is hallucinations in importing packages~\cite{Package-Hallucinations-1:journals/corr/abs-2406-10279, robustimports:latendresse2025} .  
This displays itself when the LLM imports a package that does not really exist, or invokes non-existent APIs of in a package. 
Locating dependencies and implementing the APIs 
correctly is another recognized problem~\cite{API-Bank:conf/emnlp/LiZ000YLHL23, Autofeedback:journals/corr/abs-2410-06943, APIproficiency:conf/kdd/0002GKEFTS24, APIinteractions:journals/corr/abs-2409-11703, Di-Bench:zhang-etal-2025-di}.  In~\cite{CodeBugsBenchmark:journals/corr/abs-2407-06153} they state that 
``For RWPB (real world projects) ... the proportion of runtime bugs, 
especially incorrect arguments and incorrect boundary condition checks, is significantly higher compared to existing benchmarks.''  We additionally identified
type inference and datatype conversion as a problem.   This issue is investigated in \cite{Types:journals/corr/abs-2504-09246}.

While other research has investigated the use of LLMs together with REST APIs, it has been primarily focused on how to generate test cases for  
OpenAPI specifications~\cite{RESTTESTGEN:9159077, AutomatedRestTesting:journals/corr/abs-2402-05102, barradas2025combiningtslllmautomate}. 
In contrast, our goal is to generate
code for applications that express their interfaces via REST APIs.   We are unaware of other works that explicitly address this problem.   

In this paper we propose a complexity metric for specifications, including dependency difficulty.   Along somewhat similar lines,  the authors in
\cite{CoderEval:conf/icse/YuSRZZMLLWX24} define
$6$ {\it runnable levels} to classify the scope required for a function to run successfully; e.g., from stand-alone functions to those that require knowledge of code 
in other source files.  They also propose a metric Acc@k to measure the percent of target functions that are referenced in at least one of the $k$ LLM-generated
sample codes. A similar measure, Recall@k, is defined in~\cite{Di-Bench:zhang-etal-2025-di}.  Our complexity metric includes
additional components, such as length of the specification and resource complexity.    
In~\cite{CodeBugsBenchmark:journals/corr/abs-2407-06153} the authors conclude that ``As the complexity of the benchmarks increases, 
the accuracy of LLMs tends to decline, especially when the lines of code and the number of APIs increase.''   They also find that ``Despite the large context windows 
of current LLMs, their capacity to handle lengthy and complex problems is still limited. They struggle to retain all problem details, and the extended code generation 
results in more syntactical errors like unmatched parentheses.''

\section{Conclusion}
\label{section-conclusion}
The goal of this paper is to evaluate how well LLM Agents do in generating code for microservice-based applications.   We defined a standard template for specifying such
applications, a framework for testing them, and a metric for scoring the complexity of a specification.
We ran experiments
on 8 microservices from 2 applications. Our results indicate, unsurprisingly, that the greater the complexity of the specification, the greater difficulty the LLM has in 
passing unit tests for the microservices. While stronger LLMs such as GPT-3o-mini do fairly well in generating code for specifications of 
medium difficulty, they do poorly on complex specifications that have more intricate business logic, more dependencies, and require
authentication and database integration.  We analyzed the key problems LLM Agents face in generating code for these specifications thereby suggesting future 
research directions to improve the capabilities of LLM Agents to generate microservice-based applications.   We showed that a fine-grained approach to code-generation,
generating code per request type as opposed to code for the entire service, produces more accurate code.

\bibliographystyle{acm}
\bibliography{Yellin-TOSEM.bib}

\appendix
\section{Prompts}
\label{appendix-prompts}
The following prompt is used to generate code for the given microservice.
\begin{Verbatim}[fontsize=\footnotesize] 
Below is (I) a description of the microservices for a public library 
application and (II) code generation guidelines. Generate code 
for the <microservice-name>  microservice following the guidelines.  
The code you generate should be complete and executable.
I.  <MSBA spec>
II. <code generation guidelines>
\end{Verbatim}

This prompt provides the MSBA description and the code generation guidelines. The MSBA descriptions for the public library and 
the restaurant nutrition MSBAs can be found online (see footnote~\ref{footnote-online}). The code generation guidelines are
given in Figure \ref{code-generation-guidelines}. The reflection and program regeneration prompts are presented in Figures
 \ref{reflection-prompt} and \ref{regeneration-prompt}.   The initial MSBA description, the previous code generated, and the error and test messages 
 are also supplied with these prompts.  
\begin{figure}
\begin{Verbatim}[fontsize=\footnotesize] 
Code Generation Guidelines
- Each microservice should be written in Python.
- The Python packages that need to be installed must be listed in a comment 
  at the beginning of the code.
- The code must be executable. Besides comments in the code, no other text 
  in the response.
- Each microservice should store its persistent data in a Mongo database. 
  Assume that Mongo is available at `mongodb://mongo:27017/'.
- If a microservice requires data that another microservice is responsible for, 
  it invokes a GET request on that microservice to retrieve that data when it  
  needs it. It does not replicate and store the data locally.
- Assume that each microservice will run in its own Docker container.
- For all services, each REST request should return a status code in addition
  to the JSON payload if there is one. For successful requests, the appropriate 
  2XX code is returned:  201 (successful POST), 204 (successful DELETE) and 
  200 (success on other requests).
  For unsuccessful requests, the appropriate 4XX or 5XX code is returned: 400 
  (Bad request), 404 (Not Found), 405 (Method Not Allowed), 415 (Unsupported 
  media type), and 500 (Internal Server Error).
\end{Verbatim}
\caption{Code Generation Guidelines}
\label{code-generation-guidelines}
\end{figure}

\begin{figure}[thb!]
\begin{Verbatim}[fontsize=\footnotesize]
Examine the error messages given above. For at most 5 errors, provide an 
explanation of (a) where the faulty code is located, (b) what the error 
is, and (c) how to fix it. Your explanation  should be precise so that a 
programmer can follow your instructions and correct the errors. Do not 
consider Warnings or the fact that the container is not running on a 
production server. If there is not enough information in the error  
messages, then consider which tests failed and examine the code to determine 
what changes to the code are needed. Think what may have caused the error, 
including preceding requests. For example, a GET request may return a 404 
status code if the previous POST request adding that item failed.
\end{Verbatim}
\caption{The reflection prompt}
\label{reflection-prompt}
\end{figure}

\begin{figure}[thb!]
\begin{Verbatim}[fontsize=\footnotesize]
When the Python Interpreter tried to execute this program, there were 
errors. Here is the summary and instructions on how to fix these errors: 
{reflection}. Re-generate the code and make sure to fix the errors.  
Check that you have fixed each error.
\end{Verbatim}
\caption{The code regeneration prompt}
\label{regeneration-prompt}
\end{figure}
 
\section{LLM-as-a-Judge Prompt}
\label{appendix-LLMasJudge}
The LLM-as-a-Judge prompt is:
\begin{Verbatim}[fontsize=\footnotesize]
Measure the complexity of implementing the following micro-services 
application specification on a scale 1-5 (1 is easy, 5 is hard):
<spec>
\end{Verbatim}
We used Claude Sonnet 4 to provide these ratings.   
 The ratings and comments provided by the LLM Judge for the restaurant MSBA is given in Figure \ref{figure-LLM-restaurant-app-complexity}.
The comments for the library app can be found online.   

\begin{figure}[thb!]
\begin{Verbatim}[fontsize=\footnotesize]
Looking at this Restaurant-Nutrition microservices application specification, I would 
rate the implementation complexity as **4 out of 5 (Hard)**.
Here's my analysis of the complexity factors:

**High Complexity Elements (Contributing to the 4 rating):**

1. **Multi-service Architecture**: Four interconnected microservices requiring proper 
    service-to-service communication and coordination
2. **Authentication & Authorization**: Token-based auth system with validation across 
    services, plus privileged user handling
3. **External API Integration**: Integration with Ninja API for nutritional data, 
   including error handling for external service failures
4. **Complex Business Logic**: 
   - Nutritional calculations (summing ingredient values)
   - Diet matching algorithms (comparing dish nutrition to user profiles)
   - Rating aggregation and average calculations
5. **Database Operations**: Full CRUD operations across multiple MongoDB collections 
   with proper data modeling
6. **Error Handling**: Comprehensive HTTP status code handling (401, 400, 404, 405, 
  415, 500) across all services
7. **Query Processing**: Complex filtering logic for dishes (nutritional constraints, 
  ingredient matching) and profiles

**Moderate Complexity Elements:**
- Docker containerization requirements
- REST API design and implementation
- Data validation and formatting (2 decimal places for floats)

**Why not a 5 (Extremely Hard):**
- Well-defined specifications with clear requirements
- Standard technologies (Python, MongoDB, REST APIs)
- No real-time processing or complex distributed systems patterns
- No advanced security beyond basic token auth

This is a substantial project that would require solid understanding of microservices 
architecture, database design, API integration, and comprehensive error handling, 
making it challenging but achievable for developers with intermediate to advanced skills.
\end{Verbatim}
\caption{Claude Sonnet 4 comments on the complexity of the restaurant MSBA.}
\label{figure-LLM-restaurant-app-complexity}
\end{figure}

\section{1-shot prompt}
\label{appendix-1-shot}
Figure \ref{1-shot-prompt} presents the example used in our 1-shot prompt.  For brevity, we only present the code for the POST request.
\begin{figure}[hb!]
\begin{Verbatim}[fontsize=\footnotesize]
Here is example code for a REST service that supports POST, GET, DELETE 
and PUT requests on "toys" using Python, Flask and MongoDB. Each toy is a 
JSON object containing the fields name (of type string), descr (of type string), 
age (of type integer), price (of type float), features (of type array, where  
each element of the array is a string), and id (unique string id).

from flask import Flask, jsonify, request, make_response
import pymongo
from bson import ObjectId
import os

app = Flask(__name__)
client = pymongo.MongoClient(`mongodb://mongo:27017/')
db = client[`toys_inventory']
toysColl = db[`toys']
@app.route(`/toys', methods=[`POST'])
def addToy():
    try:
        content_type = request.headers.get(`Content-Type')
        if content_type != `application/json':
            return jsonify({"error": "Expected application/json"}), 415
        data = request.get_json()
        required_fields = [`name', `age', `price']
        if not all(field in data for field in required_fields):
            return jsonify({"error": "Malformed data"}), 400
        if `features' not in data:
            features = []
        else:
            features = data[`features']
        if `descr' not in data:
            descr = "Not Available"
        else:
            descr = data[`descr']
        toy = {
            `name':data[`name'],
            `descr':descr,
            `age':data[`age'],
            `price':data[`price'],
            `features':features   
            }
        result = toysColl.insert_one(toy)
        id = str(result.inserted_id)
        response_data = {"id":id}
        return jsonify(response_data),201
    except Exception as e:
        return jsonify({"server error":str(e)}),500
\end{Verbatim}
\caption{Example used in the 1-shot prompt}
\label{1-shot-prompt}
\end{figure}

\section{Prompt for fine-grained microservice generation}
\label{appendix-dishes-POST-prompt}
The prompt to generate code {\it only} for a POST request  for the dishes microservice is given in Figure \ref{fine-grained-prompt}.   Prompts for the other request types and other 
services are similar.   
\begin{figure}[thb!]
 \begin{Verbatim}[fontsize=\footnotesize] 
Below is (I) a description of the microservices for a restaurant nutrition application 
and (II) code generation guidelines. Generate code for the POST request for the 
dishes microservice following the guidelines. When generating the code for the 
POST request, you should specify the following before the generated function 
definition:
@app.route('/dishes', methods=[`POST'])
@authenticate(required_privilege=True)

By specifying @authenticate, you need not explicitly authenticate the user.
You can assume that the generated code will be embedded in a file containing the 
following import statements:
from flask import Flask, request, jsonify
from pymongo import MongoClient
import requests
import json
import os
from functools import wraps

You can also assume that the generated code will be embedded in a file containing 
the code:
"if __name__ == `__main__':"
so do not generate another statement of this form.
I.  <MSBA spec>
II. <code generation guidelines>
\end{Verbatim}
\caption{The fine-grained prompt}
\label{fine-grained-prompt}
\end{figure}

\section{The effects of temperature}
\label{appendix-temperature} 
To see the effects of temperature, we analyze GPT-o4-mini on the $0$-shot prompt for the library MSBA.  Table 
\ref{figure-restaurant-temp-gpt-4o-mini-0-shot} presents the results.  The meaning of the columns are the same as 
in other tables of Section \ref{section-experiments}.  Each row gives the results for one temperature for one MS, and the row  {\it All}  is the average across
all temperatures for that MS\footnote{Note that the averages given for the All column cannot be obtained by simply adding up the
the percent passed for the $3$ temperatures and then dividing by $3$.  This is because to compute the average, we only consider
those experiments that actually generated testable code.   Hence to validate the results of the {\it All} row you must
take a weighted sum of the $3$ temperatures where the weights are determined by the number of testable programs 
that were generated.}.

In this table, for V0 code, temperature $0.3$ outperforms the other temperatures except for one MS (Cardholder), where temperature 
$0$ performs best. The difference between the top and second best performing temperatures are always within $10\%$, and for $2$
of the MSes (Borrows and Logs), all the temperatures are within  $10\%$ of the top performing temperature.  
For V1 code temperature $0.5$ performs best across all
MSes.  

\begin{table}
\begin{tabular}{| c | c | c |  c | c |  c | }
\hline
Name           & Temp  &  \# V0       &  \# V1       &  \# V0       &  \% V1     \\
                     &           & testable   & testable    & passed     & passed     \\
\hline
Cardholder     & 0.0    & 5/5         & 5                & 77.1         & 80.0       \\
Cardholder     & 0.3    & 4/5         & 3                & 48.2         & 52.4       \\
Cardholder     & 0.5    & 4/5         & 4                & 69.6         & 82.1       \\
Cardholder     & All     & 13/15     &12               & 65.9         & 73.8       \\
\hline
Books            & 0.0    & 1/5         & 1                & 35.7         & 35.7       \\
Books            & 0.3    & 4/5         & 2                & 73.2         & 60.7       \\
Books            & 0.5    & 4/5         & 3                & 71.4         & 73.8       \\
Books            & All     & 9/15       &6                 & 68.2         & 63.1       \\
\hline
Borrows         & 0.0    & 5/5         & 5                & 57.1         & 58.6       \\
Borrows         & 0.3    & 5/5         & 5                & 61.4         & 68.6       \\
Borrows         & 0.5    & 5/5         & 5                & 55.7         & 71.4       \\
Borrows         & All     & 15/15     & 16              & 58.1         & 66.2       \\
\hline
Logs              & 0.0    & 5/5         & 5                & 85.2         & 85.2       \\
Logs              & 0.3    & 5/5         & 1                & 94.2         & 85.7       \\
Logs              & 0.5    & 5/5         & 4                & 88.6         & 85.7       \\
Logs              & All     & 15/15     & 10              & 89.4         & 85.5       \\ 
\hline
\end{tabular}
\newline
\caption{Effects of temperature: Library MSBA, GPT-4o-mini, $0$-shot}
\label{figure-restaurant-temp-gpt-4o-mini-0-shot}
\end{table}

\end{document}